\documentclass[twoside]{article}
\usepackage[latin1]{inputenc}
\usepackage[dvips]{graphicx,epsfig,color}
\usepackage{wrapfig,rotating}
\usepackage{amssymb,amsmath,array}

\begin{document}
\title{Valence Quarks Polarization from COMPASS}

\author{\normalsize A.Korzenev\thanks{Supported by the BMBF.~ On leave from JINR, Dubna, Russia.~E-mail: korzenev@mail.cern.ch},~
  for the COMPASS collaboration
%
\vspace{.cm}\\
\vspace{.2cm}
{\normalsize Mainz University, Institute of Nuclear Physics, D-55099, Mainz, Germany}\\
\vspace{-0.1cm}
{\normalsize \it Talk given on XV Internetional Workshop on Deep-Inelastic Scattering}\\
\vspace{-0.1cm}
{\normalsize \it and Related Subjects, April 16-20, 2007, Munich, Germany}\\
\vspace{0.cm}
{\normalsize \it  \texttt{http://www.mppmu.mpg.de/dis2007}}
}

\date{}

\maketitle

\vspace{-.8cm}
\begin{abstract}
A first evaluation of the polarized valence quark distribution
$\Delta u_v(x)+\Delta d_v(x)$ from the COMPASS experiment (CERN/SPS)
is presented. The data were collected by COMPASS in the years 2002--2004
using a 160 GeV polarized muon beam scattered off a large polarized $^6$LiD
target and cover the range $1< Q^2 < 100$ GeV$^2$ and $0.006<x<0.7$.
The analysis is based on the difference asymmetry, $A^{h^+ - h^-}$,
for hadrons of opposite charges, which
gives a direct access to the polarization of valence quarks.
\end{abstract}


\section{Introduction}

Nowadays, there is growing of interest in semi-inclusive deep inelastic scattering
experiments (SIDIS) with longitudinally polarized beams and targets as they provide
an additional information on the spin structure of the nucleon compared to the
inclusive DIS measurements. 
The SIDIS data allow to separate the valence and sea contributions to the nucleon spin.

Previous measurements of the valence quark helicity distributions were done by
the SMC \cite{SMC96,SMC98} and the HERMES \cite{HERMES} collaborations.
The SMC data cover a similar kinematic range as the COMPASS data,
but with statistics which is an order of magnitude lower.
HERMES has a high statistics  data set and PID, thus it can disentangle all five
quark helicity distributions. 
However the $x$-range is quite limited: $0.023 < x < 0.6$.


In the present analysis we use the so called difference asymmetry which is determined 
from the difference of cross sections of positive and negative hadrons $h^+$ and $h^-$:
\begin{eqnarray}  
A^{h^+ - h^-} = \frac{(\sigma_{\uparrow\downarrow}^{h+}-\sigma_{\uparrow\downarrow}^{h-})-(\sigma_{\uparrow\uparrow}^{h+}-\sigma_{\uparrow\uparrow}^{h-})}{(\sigma_{\uparrow\downarrow}^{h+}-\sigma_{\uparrow\downarrow}^{h-})+(\sigma_{\uparrow\uparrow}^{h+}-\sigma_{\uparrow\uparrow}^{h-})} .
\label{def_diff_asym}  
\end{eqnarray}
Here arrows indicate the relative direction of the beam and target polarizations.
The difference asymmetry approach was developed and used in SMC \cite{Frankfurt,SMC96}.
Results obtained with this approach,
as compared to the traditional single hadron approach \cite{SMC98,HERMES},
are "cleaner" from the theoretical point of view 
because of the very weak sensitivity of $A^{h^+ - h^-}$ to uncertainties coming
from fragmentation functions (FF). 
%
%
%
As it is shown in \cite{Frankfurt} FFs cancel out 
from $A^{h^+ - h^-}$ in LO QCD. 
For the deuteron target the asymmetry is:
\begin{equation}  
A_{d}^{h^+ - h^-}\! \equiv  A_{d}^{\pi^+ - \pi^-} \! = A_d^{K^+ - K^-} \! = \frac{\Delta u_v + \Delta d_v}{u_v + d_v} \,, 
~~~~{\rm where}~~~~\Delta q_v \equiv \Delta q-\Delta \bar{q}.~
\label{diff_deuteron}  
\end{equation}
The fact that kaons contribute to the asymmetry exactly like pions allows to avoid
statistical losses due to hadron identification.
Starting from NLO QCD the difference asymmetry depends also on
FFs. However their effect is small \cite{SSI}.

The single hadron asymmetries $A^{h+}$ and $A^{h-}$ can be used to obtain $A^{h^+ - h^-}$:
\begin{eqnarray}  
A^{h^+ - h^-} = \frac{1}{1-r} ( A^{h+} - r A^{h-}) \,, & {\rm with}& 
r=\frac{\sigma_{\uparrow\downarrow}^{h-}+\sigma_{\uparrow\uparrow}^{h-}} 
{\sigma_{\uparrow\downarrow}^{h+}+\sigma_{\uparrow\uparrow}^{h+}} =  
\frac{N^-}{N^+} \cdot \frac{a^+}{a^-} .
\label{rel_diff_asym} 
\end{eqnarray}
The ratio of cross sections for negative and positive hadrons, $r$,
depends on the event kinematics and could, in principle, be measured in unpolarized
experiments. In practice, it will be obtained from the hadron number ratio $N^-\!/N^+$
corrected by the ratio of their acceptances $a^-\!/a^+$. 


Since the deuteron is an isoscalar target we can not distinguish between  
up and down quarks. 
Nevertheless having measured the first moment of $\Delta u_v(x)$$+$$\Delta d_v(x)$
and combining its value with axial charges $a_0$ and $a_8$
the information about the symmetry of sea quark distributions can be extracted. 
Since \,$\Delta s+\Delta\bar{s}=\frac{1}{3}(a_0-a_8)$ one can show that
\begin{eqnarray}  
\Delta \bar{u} + \Delta \bar{d} 
 = (\Delta s + \Delta \bar{s}) + \frac{1}{2}(a_8 - \Gamma_v),
~~~{\rm where}~~~\Gamma_v = \int_0^1 (\Delta u_v(x) + \Delta d_v(x)) dx .~~
\label{Eq:main_eq} 
\end{eqnarray} 
%
The $SU(3)_f$ symmetric sea ($\Delta\bar{u}=\Delta\bar{d}=\Delta s=\Delta\bar{s}$) 
will obviously lead to $\Gamma_v$$=$$a_8$. 
In contrast, if measurements give $\Gamma_v=a_8+2(\Delta s + \Delta\bar{s})$  
it will point to a strong asymmetry for the first moments of light sea quarks 
$\Delta\bar{u}=-\Delta\bar{d}$. 

\begin{figure}
\includegraphics[width=0.49\columnwidth]{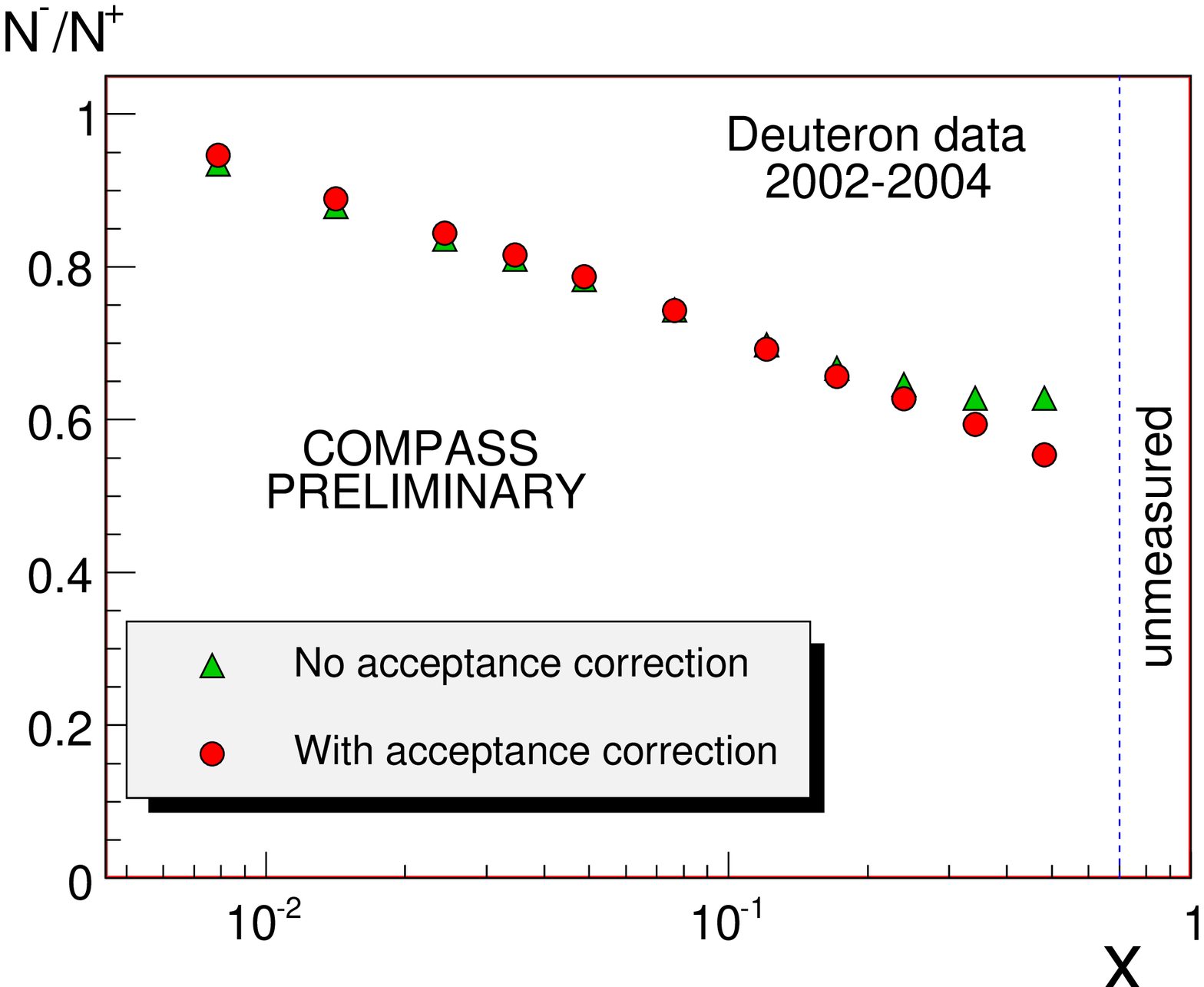}
\hfill
\includegraphics[width=0.49\columnwidth]{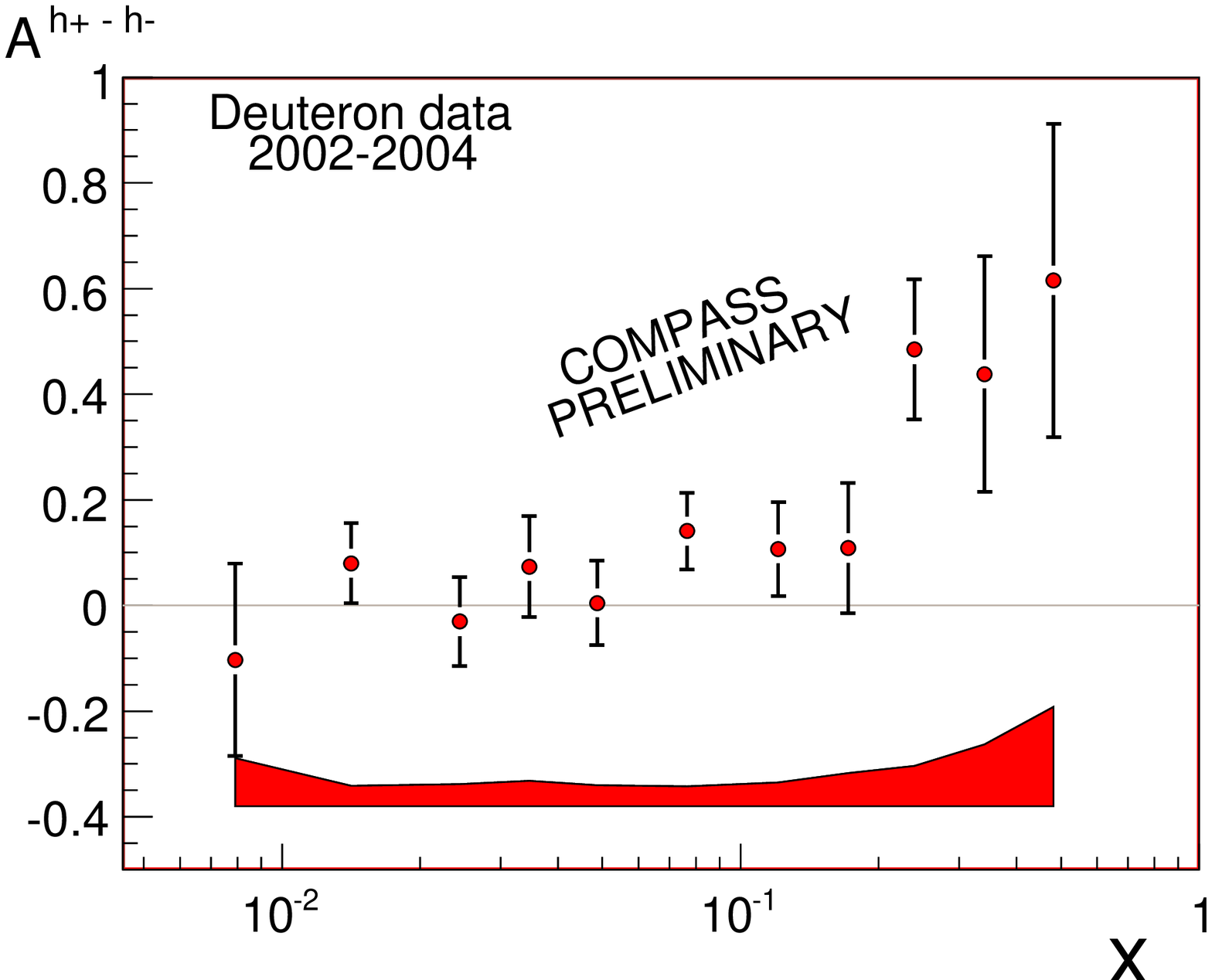}
\caption{Left: The hadron number ratio $N^{-}\!/N^{+}$  and  
    the same ratio corrected for the acceptance  which 
    represents $\sigma^{h^{-}}\!/\sigma^{h^{+}}$.
  Right: Difference asymmetry $A^{h^+-h^-}$.
}
\label{Fig:1}
\end{figure}

\section{Extraction of the asymmetry}

In the analysis data collected during the years 2002--2004 have been used.
We require for all events to have a reconstructed primary interaction
vertex defined by the incoming and the scattered muons.
The energy of the beam muon is constrained to be in the interval \mbox{$140<E_\mu<180$ GeV}.
To equalize fluxes through the two target cells it is required for
the trajectory of the incoming muon to cross both cells.
The kinematic region is defined by cuts on the photon virtuality $Q^2$
and the fractional energy $y$ transfered from the beam muon to the virtual photon.
The requirement $Q^2>1$\,GeV$^2$ selects the region of DIS.
The cut $y>0.1$ removes events which are problematic from reconstruction 
point of view due to a small energy transfer. 
The region which is the most affected by radiative corrections
is eliminated with the cut $y<0.9$.
At low $x$ (high $W$), where cross-sections of positive and negative
hadrons are approximately equal, the statistical error of $A^{h^+ - h^-}$
increases drastically.
Due to this reason we consider only $x > 0.006$.
For hadron tracks coming from the primary vertex the cut $z>0.2$
is applied to select the current fragmentation region. 
Hadron identification is not used.
The resulting sample contains 30 and 25 millions of positive 
and negative hadrons, respectively.


The contributions to the systematic error from the target
and beam polarizations, the dilution and depolarization
factors amount to 8\% of the asymmetry value when added in quadrature.
The upper limit of the false asymmetry which could be generated by instabilities 
of the spectrometer components was evaluated as a fraction of statistical error:
$\sigma_{false}<0.5 \sigma_{stat}$.
The asymmetry $A^{h^+ - h^-}$ with its statistical and systematic errors 
is shown in Fig.\,\ref{Fig:1}.


\begin{figure}
\includegraphics[width=0.49\columnwidth]{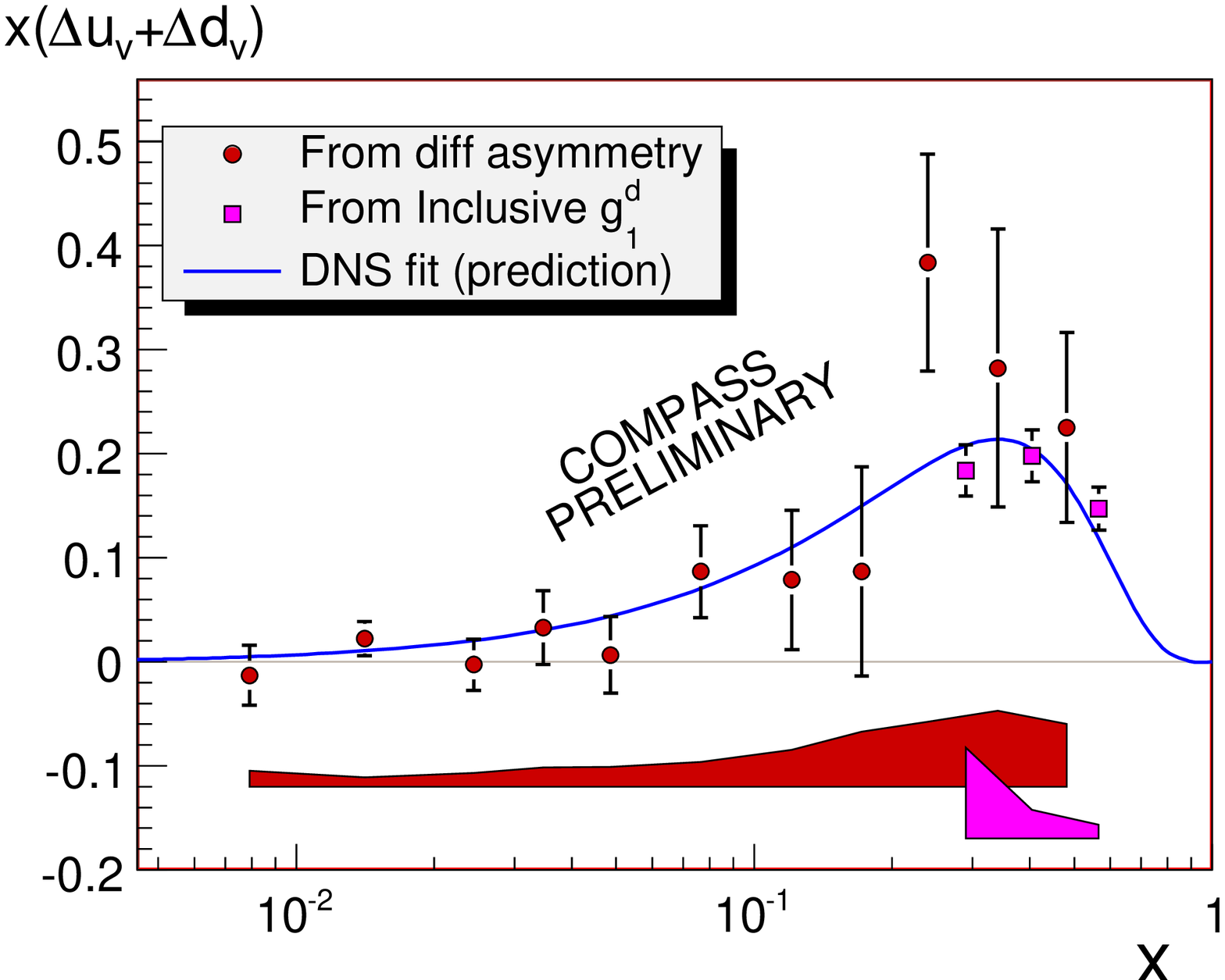}
\hfill
\includegraphics[width=0.49\columnwidth]{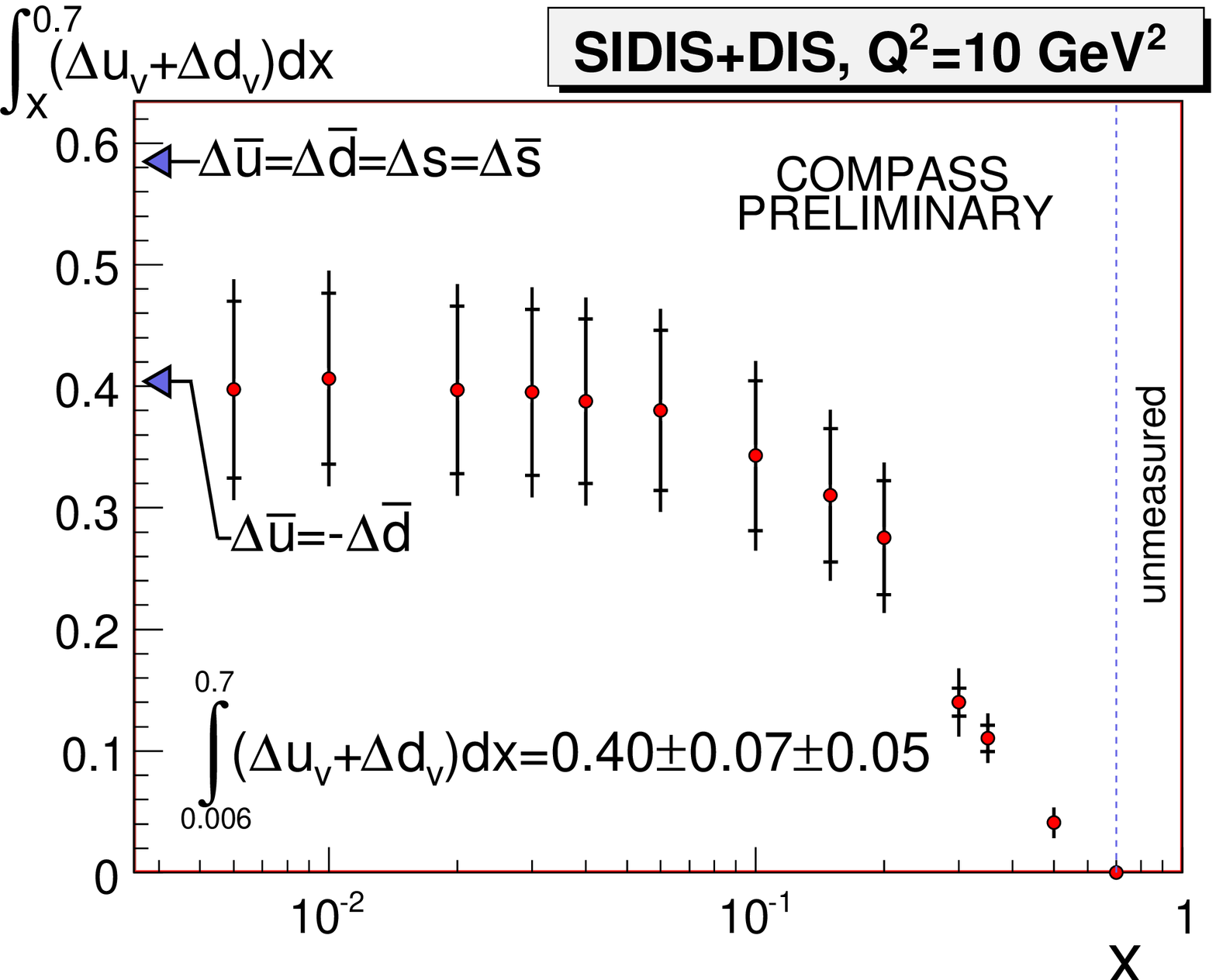}
\caption{Left: Polarized valence quark distribution $x(\Delta u_v$$+$$\Delta d_v)$ 
  evolved to $Q^2 = 10$\,GeV$^2$ according to the DNS fit at LO \cite{dns}. 
  The line shows the prediction from the fit.  
  Right: Corresponding integral of $\Delta u_v(x)$$+$$\Delta d_v(x)$ 
  as the function of the low $x$ limit of integration.
}
\label{Fig:2}
\end{figure}

\section{Extraction of $\Delta u_v$$+$$\Delta d_v$  and its first moment}

The values of $\Delta u_v$$+$$\Delta d_v$ are
obtained by multiplying $A^{h^+ - h^-}$ by the unpolarized valence distribution of
MRST\,2004 at LO \cite{mrst}.
Corrections for the deuteron D-state contribution and for the fact that 
the unpolarized parton distributions originates from $F_2$, in which 
$R = \sigma_L/\sigma_T$ was different from zero, are applied
\begin{equation}
\Delta u_v + \Delta d_v = \frac{u_v + d_v}{(1 + R(x,Q^2)) (1 - 1.5 \omega_D)} A^{h^+ - h^-}. 
\label{eq:SIDIS_Apm} 
\end{equation}

The evaluation of the first moment, $\Gamma_v$, requires the evolution of 
all $\Delta u_v(x)$$+$$\Delta d_v(x)$ points to a common $Q^2$.
This is done by using the DNS parametrization in LO \cite{dns}
which is based on the global QCD analysis of all DIS $g_1$ data prior to 
COMPASS as well as the SIDIS data from SMC and HERMES.
The parametrization corresponding to KKP fragmentation functions was used.
The resulting distribution at $Q^2$=10\,GeV$^2$ is shown in Fig.\,\ref{Fig:2}.
A good agreement of the curve with the COMPASS points 
illustrates the consistency between the three experiments.

For $x>0.3$ the unpolarized sea contribution to $F_2$ practically vanishes.  
Due to positivity conditions $|\Delta q|<q$
the polarized sea contribution to the spin of the nucleon also can be neglected. 
It allows to replace at LO Eq.\,(\ref{eq:SIDIS_Apm}) by 
\begin{equation} 
  \Delta u_v + \Delta d_v ~=~ \frac{36}{5} \frac{g_1^d(x,Q^2)}{(1 - 1.5 \omega_D)} 
  - 
  \Bigl( 2(\Delta\bar{u}+\Delta\bar{d})+\frac{2}{5}(\Delta s+\Delta\bar{s}) \Bigr) 
  \label{eq:DIS_g1d} 
\end{equation} 
which gives a much more precise evaluation of $\Delta u_v + \Delta d_v$ at high $x$. 
In the calculation we omit the second term of the right side of this equation. 
However it is used to evaluate the systematic error. 
The values of $g_1^d$ from \cite{g1d} were used.
In total, we obtain 
\begin{equation} 
  \Gamma_v (0.006<x<0.7)\Bigl|_{{\rm Q}^2=10\, {\rm GeV}^2} = 
  0.40 \pm 0.07 ({\rm stat.}) \pm 0.05 ({\rm syst.}), 
  \label{eq:Gammav_DIS_SIDIS} 
\end{equation} 
which is $2\sigma$ below the value corresponding to a flavor symmetric sea and 
very close to the value expected for $\Delta {\overline u} =\! - \Delta {\overline d}$
(see Eq.\,(\ref{Eq:main_eq}) where $\Delta s$$+$$\Delta\bar{s}$ is taken from \cite{g1d}). 
The comparison with first moments obtained with results of SMC and HERMES 
can be found in Tab.\,\ref{tab:FirstMom}.


As one can judge from Fig.\,\ref{Fig:2} the integral is practically
constant at low $x$. Thus the low $x$ contribution to $\Gamma_v$ is expected to be negligible.
The contribution to $\Gamma_v$ for $x > 0.7$ estimated with the LO DNS  
parametrization is 0.004.

\begin{table}
\begin{tabular}{|l||c|c||c|c|c|c|}
\hline
 & $x$-range & $Q^2$ & \multicolumn{2}{c|}{~$\Delta u_v + \Delta d_v$~} & \multicolumn{2}{c|}{~$\Delta\bar{u} + \Delta\bar{d}$~} \\
\cline{4-7}
   &  &  \!\!GeV$^2$\!\!\! & \!Value of the exper.\!\! & DNS   &  \!Value of the exper.\!\!  & DNS \\
\hline
\hline
\!\!SMC\,98   & 0.003--0.7 & 10  & \!\!$0.26\pm0.21\pm0.11$\!\! & 0.386 & ~$0.02\pm0.08\pm0.06$\!\! &  \!$-0.009$\! \\
\hline
\!\!HERMES\,05\!\!& 0.023--0.6 & 2.5 & \!\!$0.43\pm0.07\pm0.06$\!\! & 0.363 & \!\!\!$-0.06\pm0.04\pm0.03$\!\! & \!$-0.005$\! \\
\hline
\hline
\!\!COMPASS & 0.006--0.7 & 10  & \!\!$0.40\pm0.07\pm0.05$\!\!  & 0.385 & ~$0.0\pm0.04\pm0.03$\!\! & \!$-0.007$\! \\
\hline
\end{tabular}
\caption{Estimates of the first moments $\Delta u_v + \Delta d_v$  
  and $\Delta\bar{u} + \Delta\bar{d}$ from the SMC \cite{SMC98}, HERMES \cite{HERMES}, 
  COMPASS data and also from the DNS fit at LO \cite{dns}. 
}
\label{tab:FirstMom}
\end{table}

\section{Conclusion}

A first LO evaluation of the polarized valence quark distribution
$\Delta u_v(x)$$+$$\Delta d_v(x)$ from the COMPASS deuteron data
is presented. The data were collected by COMPASS in the years 2002--2004
and cover the range $1< Q^2 < 100$ GeV$^2$ and $0.006<x<0.7$.
The analysis was based on the difference asymmetry approach.
It leads to the first moment of $\Delta u_v$$+$$\Delta d_v$:  $0.40 \pm 0.07 ({\rm stat.}) \pm 0.05 ({\rm syst.})$ which favors the ``asymmetric'' light sea scenario
$\Delta {\overline u} = - \Delta {\overline d}$
as compared to the ``symmetric'' one $\Delta {\overline u} = \Delta {\overline d}= \Delta s = \Delta {\overline s}$. 


\begin{footnotesize}



%

\end{footnotesize}


\end{document}